\documentstyle[preprint,aps]{revtex}

\begin{document}
\draft
\title{Submergence of the Sidebands in the Photon-assisted Tunneling through
a Quantum Dot Weakly Coupled to Luttinger Liquid Leads}
\author{{\ Yi-feng Yang and Tsung-han Lin}$^{\ast }$}
\address{{\it State Key Laboratory for Mesoscopic Physics and }\\
{\it Department of Physics, Peking University,}{\small \ }{\it Beijing}\\
100871, China}
\maketitle

\begin{abstract}
We study theoretically the photon-assisted tunneling through a quantum dot
weakly coupled to Luttinger liquids (LL) leads, and find that the zero bias
dc conductance is strongly affected by the interactions in the LL leads. In
comparison with the system with Fermi liquid (FL) leads, the sideband peaks
of the dc conductance become blurring for $\frac{1}{2}<g<1,$ and finally
merge into the central peak for $g<\frac{1}{2}$ ($g$ is the interaction
parameter in the LL leads). The sidebands are suppressed for LL leads with
Coulomb interactions strong enough, and the conductance always appears as a
single peak for any strength and frequency of the external time-dependent
field. Furthermore, the quenching effect of the central peak for the FL case
does not exist for $g<\frac{1}{2}.$
\end{abstract}



\baselineskip 20pt 
\newpage

The photon-assisted tunneling (PAT) through a quantum dot weakly coupled to
two Fermi liquid leads (FL-QD-FL) has been studied intensively in the last
decade.\cite{kouwen,blick,ooster} When two time-dependent fields with the
same form $V\cos (\Omega t)$ are applied on the leads symmetrically, a
series of additional peaks (called sidebands) occurs in the curve of the
zero bias dc conductance vs. the gate voltage. These peaks locate at
energies $\delta E=n\hbar \Omega $ away from the original resonant level.
The $n$th sideband will be quenched as $\alpha $ equal to zeros of the
Bessel function $J_{n}$,\cite{wagner} because the height of the $n$th peak
is proportional to $J_{n}^{2}\left( \alpha \right) $ (where $\alpha =\frac{eV%
}{\hbar \Omega }$ is a dimensionless variable). Physically, all these
features of PAT can be ascribed to characters of the electron occupations in
the leads, or equivalently the characters of the density of states and the
Fermi distribution of the electrons in the leads.

One dimensional interacting electron systems behave as Luttinger liquid (LL),%
\cite{tomo,lutt,hald} for that theory predicts the power-law behavior of the
density of states (DOS) of the electrons.\cite{voit} What will happen for
the sidebands features if the two FL leads are replaced by two LL leads? We
expect that these features may be strongly affected by the specific
characters of the DOS in LL leads. With this idea in mind, we study in this
letter the PAT through a QD weakly coupled to two LL leads, i.e., a LL-QD-LL
system.

In contrary with the FL-QD-FL system, we find that the sideband peaks of the
dc conductance become blurring for $\frac{1}{2}<g<1,$ and finally merge into
the central peak for $g<\frac{1}{2}$. The sidebands are suppressed for LL
leads with Coulomb interactions strong enough, and the conductance appears
as a single peak for any strength and frequency of the external
time-dependent field. We call this {\it the submergence of the sidebands}.
The contributions of each PAT process to the corresponding sidebands are
diminished by the suppression of the DOS near Fermi energy, and shared due
to the unusual DOS of the LL leads. Furthermore, in strong interaction case $%
\left( g<\frac{1}{2}\right) $, the central peak consists not only of the
direct tunneling process but also many other type PAT processes, so that it
does not exist a single parameter $\alpha $ by which the central peak can be
quenched.

We use the nonequilibrium-Green-function technique to calculate the average
current through the quantum dot.\cite{haug} Under the adiabatic
approximation, the electron occupations in each lead remains unchanged. Thus
the exact time-dependent Green functions in the uncoupled leads are $\qquad $%
\begin{equation}
g_{L/R}^{r,a,<,>}(t_{1},t_{2})=\exp (-i\int_{t_{2}}^{t_{1}}\Delta
_{L/R}(t)dt)g_{0,L/R}^{r,a,<,>}(t_{1},t_{2}),
\end{equation}%
where $g_{0,L/R}^{r,a,<,>}(t_{1},t_{2})$ are the Green functions of the
uncoupled leads without the time-dependent field, and $\Delta
_{L/R}(t)=eV_{L/R}\cos (\Omega t)$. We take the units of $\hbar =1$
hereafter.

The Hamiltonian of the system can be split into three parts: $%
H=H_{leads}+H_{d}+H_{T}$, where $H_{leads}=H_{L}+H_{R}$ denotes the left and
right LL leads, $H_{d}=\varepsilon d^{+}d$ is the Hamiltonian of the QD,
with $\{d^{+},d\}$ the creation/annihilation operators of the electron in
the dot and only one energy level is considered. $H_{T}$ is the tunneling
Hamiltonian and can be written as:

\begin{equation}
H_{T}=\sum_{\lambda =L,R}(t_{\lambda }d^{+}\psi _{\lambda }+h.c),
\end{equation}%
in which $\{\psi _{\lambda }^{+},\psi _{\lambda }\}(\lambda =L/R)$ are the
Fermi operators at the end points of the left/right lead; $\{t_{\lambda }\}$
is the tunneling constant. The spin indices have been suppressed since they
are not important in the following discussion.

Using nonequilibrium-Green-function technique, the current through the
quantum dot can be formulated as:\cite{yang}

\begin{equation}
J(t)=J_{L}(t)=-2e%
\mathop{\rm Re}%
\int dt_{1}\left[ t_{L}^{\ast }\left( G^{r}\left( t,t_{1}\right)
g_{L}^{<}\left( t_{1},t\right) +G^{<}\left( t,t_{1}\right) g_{L}^{a}\left(
t_{1},t\right) \right) t_{L}\right] ,
\end{equation}%
where we have defined $G^{r}\left( t,t_{1}\right) =-i\theta
(t-t_{1})\left\langle \{d\left( t\right) ,d^{+}\left( t_{1}\right)
\}\right\rangle _{H}$ and all the other Green functions with the standard
forms.\cite{haug}

Supposing $J(t)=\sum_{n}J_{n}\exp (in\Omega t)$, then

\begin{equation}
J_{n}=-2e\left| t_{L}\right| ^{2}%
\mathop{\rm Re}%
\sum_{m}\int \frac{d\omega }{2\pi }\left( G_{m,0}^{r}\left( \omega \right)
g_{L,n,m}^{<}\left( \omega \right) +G_{m,0}^{<}\left( \omega \right)
g_{L,n,m}^{a}(\omega \right) ,
\end{equation}
in which the functions $F_{n,m}(\omega )\equiv F_{n-m}(\omega +m\Omega ),$
and $F_{n}(\omega )$ is the double Fourier transformation of function $%
F(t,t^{^{\prime }})$:\cite{sun}

\begin{equation}
F(t,t^{^{\prime }})=\sum_{n}\exp (in\Omega t^{^{\prime }})\int \frac{d\omega 
}{2\pi }e^{-i\omega \left( t-t^{^{\prime }}\right) }F_{n}(\omega ).
\end{equation}

The retarded Green function of the QD in equation $\left( 4\right) $, $%
G^{r}\left( t,t_{1}\right) $, can be calculated by using Dyson equation

\begin{equation}
G^{r}\left( t,t_{1}\right) =g^{r}\left( t,t_{1}\right) +\int \int d\tau
d\tau ^{^{\prime }}g^{r}\left( t,\tau \right) \Sigma ^{r}\left( \tau ,\tau
^{^{\prime }}\right) G^{r}\left( \tau ^{^{\prime }},t_{1}\right) ,
\end{equation}%
where $\Sigma ^{r/a}\left( \tau ,\tau ^{^{\prime }}\right) =\sum_{\lambda
=L,R}\left| t_{\lambda }\right| ^{2}g_{\lambda }^{r/a}\left( \tau ,\tau
^{^{\prime }}\right) $ is the irreducible self-energy, $g^{r}\left(
t,t_{1}\right) $ is the free retarded Green function of isolated dot. Notice
that the Green functions of the QD depend on two time variables, not the
time difference; so one should take the double Fourier transform, and obtain:

\begin{equation}
G_{n,m}^{r}(\omega )=g_{n,m}^{r}(\omega
)+\sum_{l_{1},l_{2}}g_{n,l_{1}}^{r}(\omega )\Sigma _{l_{1},l_{2}}^{r}(\omega
)G_{l_{2},m}^{r}(\omega ).
\end{equation}

Neglecting the off-diagonal elements of $\Sigma _{l_{1},l_{2}}^{r}(\omega )$%
, we have

\begin{equation}
G_{n,m}^{r}(\omega )=\frac{\delta _{n,m}}{\omega -\varepsilon +n\Omega
-\sum_{l,\lambda }J_{l+m}^{2}(\alpha _{\lambda })\left| t_{\lambda }\right|
^{2}g_{0,\lambda }^{r}(\omega -l\Omega )},
\end{equation}%
in which $J_{l}(\alpha )$ is the Bessel function, $\varepsilon $ is the
energy level in the dot and $\alpha _{L/R}=\frac{eV_{L/R}}{\Omega }.$ The
lesser Green function can be calculated using Keldysh formalism:\cite{haug}

\begin{equation}
G_{n,m}^{<}(\omega )=\sum_{l_{1},l_{2}}G_{n,l_{1}}^{r}(\omega )\Sigma
_{l_{1},l_{2}}^{<}(\omega )G_{l_{2},m}^{a}(\omega ).
\end{equation}

Substituting equation $\left( 8\right) $ and $\left( 9\right) $ into
equation (4), we can obtain the average current as:

\begin{equation}
\left\langle J\right\rangle =e\left| t_{L}t_{R}\right| ^{2}\int \frac{%
d\omega }{2\pi }\frac{(\sum_{m}J_{m}^{2}(\alpha _{L})g_{0,L}^{>}(\omega
-m\Omega ))(\sum_{m}J_{m}^{2}(\alpha _{R})g_{0,R}^{<}(\omega -m\Omega
))-(L\leftrightarrow R)}{(\omega -\varepsilon )^{2}+\left| (\sum_{m,\lambda
}J_{m}^{2}(\alpha _{\lambda })g_{0,\lambda }^{r}(\omega -m\Omega ))\right|
^{2}}.
\end{equation}

Equation (10) is the main result of our work. For the model discussed in
this letter, the Green functions $g_{0,L/R}^{<,>}$ are taken from \cite{furu}%
:

\begin{eqnarray}
g_{0,L}^{<,>}\left( \varepsilon \right) &=&\pm i\frac{T}{\left| t_{L}\right|
^{2}}e^{\mp \frac{\varepsilon }{2T}}\gamma _{L}\left( \varepsilon \right) ,
\\
g_{0,R}^{<,>}\left( \varepsilon \right) &=&\pm i\frac{T}{\left| t_{R}\right|
^{2}}e^{\mp \frac{\varepsilon -eV}{2T}}\gamma _{R}\left( \varepsilon
-eV_{b}\right) ,  \nonumber
\end{eqnarray}%
where $T$ is the temperature, and $V_{b}$ the dc bias voltage. $\gamma
_{L/R}\left( \varepsilon \right) $ is defined as $\gamma _{L/R}\left(
\varepsilon \right) =\frac{\Gamma _{L/R}}{2\pi \left| t_{L/R}\right| ^{2}}%
\left( \frac{\pi T}{\Lambda }\right) ^{1/g_{L/R}-1}\frac{\left| \Gamma
(1/2g_{L/R}+i\varepsilon /2\pi T)\right| ^{2}}{\Gamma \left(
1/g_{L/R}\right) }$, here $\Gamma \left( x\right) $ is the Gamma function, $%
g_{L/R\text{ }}$ is the interaction parameters characterizing the left/right
LL liquids, $\Gamma _{L/R}$ describes the effective level broadening of the
dot, proportional to $\left| t_{L/R}\right| ^{2}$ , and $\Lambda $ is the
high-energy cutoff or a band width. The Hamiltonian can describe a quantum
dot coupled to the end points of either two half-infinite quantum wires or
two edge states of fractional quantum Hall liquids with filling factor $\nu =%
\frac{1}{2m+1}=g.$ The average current formula $\left( \text{Eq.}\left(
10\right) \right) $ covers a wide range of interaction in the leads, from
noninteracting limit $\left( g=1\right) $ to strong interacting LL $\left( 
\text{small }g\right) .$

What we are interested in is the dc conductance at zero bias. We consider
the symmetry case, namely $t_{L}=t_{R}=t,g_{L}=g_{R}=g,$ and $\alpha
_{L}=\alpha _{R}=\alpha $. If the external fields are not very strong,
equation $\left( 10\right) $ can be further simplified. Under the condition $%
\frac{\Gamma }{T}\left( \frac{\varepsilon }{2\Lambda }\right) ^{1/g-1}\ln
\left( \frac{\varepsilon }{2T}\right) \ll 1(T\ll \frac{\varepsilon }{2})$ or 
$\frac{\Gamma }{\Lambda }\left( \frac{T}{\Lambda }\right) ^{1/g-2}\ll 1(T\gg 
\frac{\varepsilon }{2}),$\cite{yang} we obtain the zero bias conductance
after a contour integral (all the subscript $L/R$ are omitted):

\begin{equation}
G=e^{2}\left| t\right| ^{2}\frac{(\sum_{m}J_{m}^{2}(\alpha
)g_{0}^{>}(\varepsilon -m\Omega ))_{\varepsilon }^{^{\prime
}}(\sum_{m}J_{m}^{2}(\alpha )g_{0}^{<}(\varepsilon -m\Omega
))-(<\leftrightarrow >)}{\left| (\sum_{m,\lambda }J_{m}^{2}(\alpha _{\lambda
})g_{0,\lambda }^{<}(\varepsilon -m\Omega ))-(\sum_{m,\lambda
}J_{m}^{2}(\alpha _{\lambda })g_{0,\lambda }^{>}(\varepsilon -m\Omega
))\right| }
\end{equation}%
\bigskip\ \ \ \ \ \ 

In the following, we use current formula (12) to study the zero bias
conductance vs. the energy level of the QD, $G/G_{\max }$ vs. $\frac{%
\varepsilon }{T},$ numerically. We consider the whole range of interaction
of the leads, from noninteracting limit ($g=1$) to strong interaction case ($%
g<\frac{1}{2}$), with much attention on the latter. Fig.1 presents the
results for $\frac{\Omega }{T}=5$, $\alpha =\frac{eV}{\Omega }=1\ $and $%
g=1.0,0.8,0.5,0.2$. The sideband peaks appears clearly for $g=1$; become
blurring for $g=0.8;$ merge into a single central peak for $g=0.5;$ and
finally become narrower for $g=0.2$.

In order to study the quenching behavior, we chose $\alpha $ equal to the
smallest positive root of the zero order Bessel function, $J_{0}$. For the
noninteracting limit case ($g=1)$, the central peak is completely quenched
as expected. However, with the increase of the interaction or the decrease
of $g,$ the situation changes significantly. For $g=0.8$, the relative
heights of the sideband peaks are almost unchanged, but the valleys lifted.
Both are shown in Fig.2(a). Most surprisingly, the quenching effect
disappears completely for the strong interaction case (shown in Fig.2(b)): a
broader peak with a flat plateau for $g=0.5$; and a narrower single central
peak for $g=0.2$.

To further understand the PAT in LL-QD-LL for strong interaction case, we
calculate $G/G_{\max }$ vs. $\frac{\varepsilon }{T}$ for $g=0.2,$ with
different strengths and frequencies of the external fields, shown in Fig.3.
One can clearly see that all sideband peaks disappear. And the width of the
single central peak becomes broader with the increase of the strength or the
frequency of the fields, originated from the contributions of the
multi-photon-assisted tunneling processes.

All features mentioned above can be ascribed to the power-law behavior of
the DOS and the suppression of the electron tunneling near the Fermi
energies of LL leads. In order to understand the underlying physics, we
label $\left( m,n\right) $ to indicate two pairs of processes: (1)the
electron tunneling through the dot with $m$ photons absorbed in the left and 
$n$ photons emitted in the right lead, and its reversal process; (2) same
processes as in (1) except $m$ $\rightleftharpoons n$. As an analogy of the $%
\left( m,n\right) ,$ one can consider the tunneling currents under two
different finite bias voltages $(m\Omega +eV_{b})-n\Omega $ and $\left(
n\Omega +eV_{b}\right) -m\Omega ,$ without the time-dependent fields. The
difference of these two net currents is approximately equal to the net
current of $\left( m,n\right) $, except the effect due to difference of the
width of the energy level in QD. In the following, we shall discuss the PAT
qualitatively with the help of this analogy.

It is well-known that for FL-QD-FL, the current under finite bias change
abruptly at the Fermi energies in the two leads due to the abrupt change of
the electron occupations near Fermi energies in the leads; thus two sharp
peaks at $\varepsilon =m\Omega $ and $n\Omega $ obtained from the difference
of the currents under two finite bias voltages, provided $V_{b}$ is small.
Similarly, the conductance from $(m,n)$ has the same peaks at $\varepsilon
=m\Omega $ and $n\Omega .$ Conductance at other energies is strongly
suppressed due to the uniform distribution of the electrons.

However, for LL-QD-LL, the DOS of the leads behaves as $\rho \left( \omega
\right) \propto \left| \omega \right| ^{1/g-1}$ as $T\rightarrow 0$, the
so-called power-law behavior of the DOS, so $\rho ^{\prime }\left( \omega
\right) \propto \left| \omega \right| ^{1/g-2}.$ For the weak interaction ( $%
g>\frac{1}{2}$ ), $\rho ^{\prime }\left( \omega \right) $ $\rightarrow
\infty $ as $\omega \rightarrow 0$, thus there is also an abrupt change in
the electron occupation near Fermi energy; two peaks at $\varepsilon
=n\Omega $ and $m\Omega $ still remain, but the valley between them is
lifted due to the nonuniform of the DOS. The contributions from all
different $\left( m,n\right) $ processes give the total conductance with
sideband peaks at $\varepsilon =\pm \Omega ,\pm 2\Omega ...,$ and the
valleys are lifted with the decrease of $g$, consistent with the curves for $%
g=1,0.8$ in Fig.1 and Fig.2$\left( a\right) .$

For the strong interaction case ( $g<\frac{1}{2}$ ), $\rho ^{\prime }\left(
\omega \right) $ $\rightarrow 0$ as $\omega \rightarrow 0$, thus no abrupt
change in the electron occupation exists, and the difference of the currents
under two finite bias voltages rises slowly at the outset and finally
appears as a single peak centered at $\frac{m+n}{2}\Omega $ approximately,
so that the $\left( m,n\right) $ processes do not contribute significantly
at $\varepsilon =m\Omega $ and $n\Omega $, instead, a peak at $\varepsilon =$
$\frac{m+n}{2}\Omega $ is produced. Furthermore, our calculation shows that
processes such as $\left( m,m\right) $, always producing a single peak at $%
\varepsilon =m\Omega $ for any values of $g$, are not important, due to the
fact that the direct tunneling near Fermi energy is strongly suppressed for
strong interaction case. In Fig.4, we present the results from processes $%
\left( 1,1\right) $ and $\left( 2,2\right) ,$ showing a contribution almost
6 orders of magnitude smaller than the main processes. Moreover, one can see
from Fig.4 that the main contributions to the conductance are originated
from processes $\left( m,-n\right) $, with $m,n=0,1,2...$, especially from
that of $m=n$, which make significant contribution at $\varepsilon =0$ but
little at the sideband energies$.$ That is why for the strong interaction
case only one single peak at $\varepsilon =0$ exists, while all the
sidebands are suppressed completely, as shown in Fig.3.

Now let us explain why the quenching of the central peak disappears for
strong interaction case. Notice that the main contributions of the central
peak come mainly from processes $\left( 0,m\right) $ with $m\in Z$ for weak
interaction ($g>1/2);$ but $\left( m,-n\right) $ with $m\approx n$ for
strong interaction case ( $g<1/2)$. Given $\alpha $ the zeros of $J_{0}$,
all the processes like $\left( 0,m\right) $ become zero and contribute
nothing to the central peak for weak interaction case, but the processes in
the strong interaction case are not affected much. Thus the central peak is
quenched as usual in the former, but will be lack in the latter $\left( 
\text{see Fig.2}\right) $. Furthermore, since the conductance at $%
\varepsilon =0$ come from many PAT processes, it does not exist a single
parameter $\alpha $ by which all the processes,having main contributions to
the central peak,can be neglected; thus there is no quenching of the central
peak in case of LL leads with $g<\frac{1}{2}.$

Formula (10) can also be applied to the case with asymmetric external
fields. If only one lead is applied with an ac field, we find that the
photon-electron pumping effect still occurs for LL case. However, the
negative current becomes smaller and smaller with the decrease of $g$ and
the maximum of the current shifts towards the negative $\varepsilon $ (not
shown here).

In conclusion, we have studied the PAT in the system of LL-QD-LL for the
whole range of interaction of the LL leads and the crossover behavior. We
find that the zero bias conductance is strongly renormalized by the
interaction. In particular, for the strong interaction case( $g<\frac{1}{2}$
), all sidebands merge into a single peak (we called it submergence of the
sidebands), and the quenching effect also disappears completely. These novel
features of PAT have been ascribed to the power-law behavior of the DOS and
the suppression of the electron tunneling near the Fermi energies of LL
leads. Since recent experiments have confirmed that the carbon nanotubes
(CNT) do have some properties of the LL,\cite{marc,egger} especially the
power-law feature of the DOS, we suggest a setup consisted of CNT-QD-CNT to
study the PAT of the system. We hope our theoretical predictions can be
tested experimentally.

This work was supported by NSFC under grant No.10074001. One of the authors
(T.H.Lin) would also like to thank the support from the State Key Laboratory
for Mesoscopic Physics in Peking University.

\smallskip $^{\ast }$ To whom correspondence should be addressed.

$
\bigskip $

\section*{Figure Captions}

\begin{itemize}
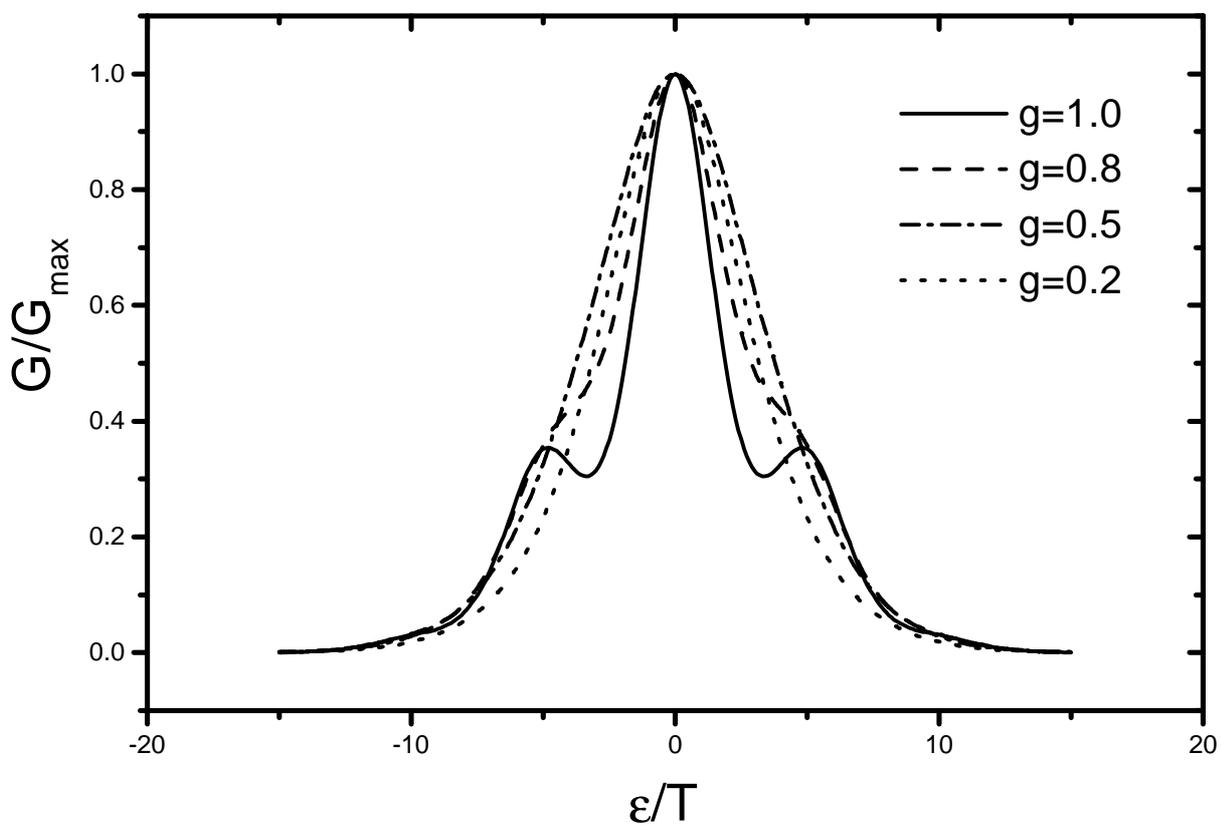
\item[{\bf Fig. 1}] $G/G_{\max }$ vs. $\varepsilon /T$ for $g=1.0,0.8,0.5,0.2
$. $G$ is the zero bias conductance, $G_{\max }$ is the maximum of $G$, $%
\varepsilon $ is the energy level in the QD, $T$ is the temperature, and $g$
is the interaction parameters of the leads. The parameters of the
time-dependent fields are: $\Omega /T=5$ and $\alpha =\frac{eV}{\Omega }=1$.

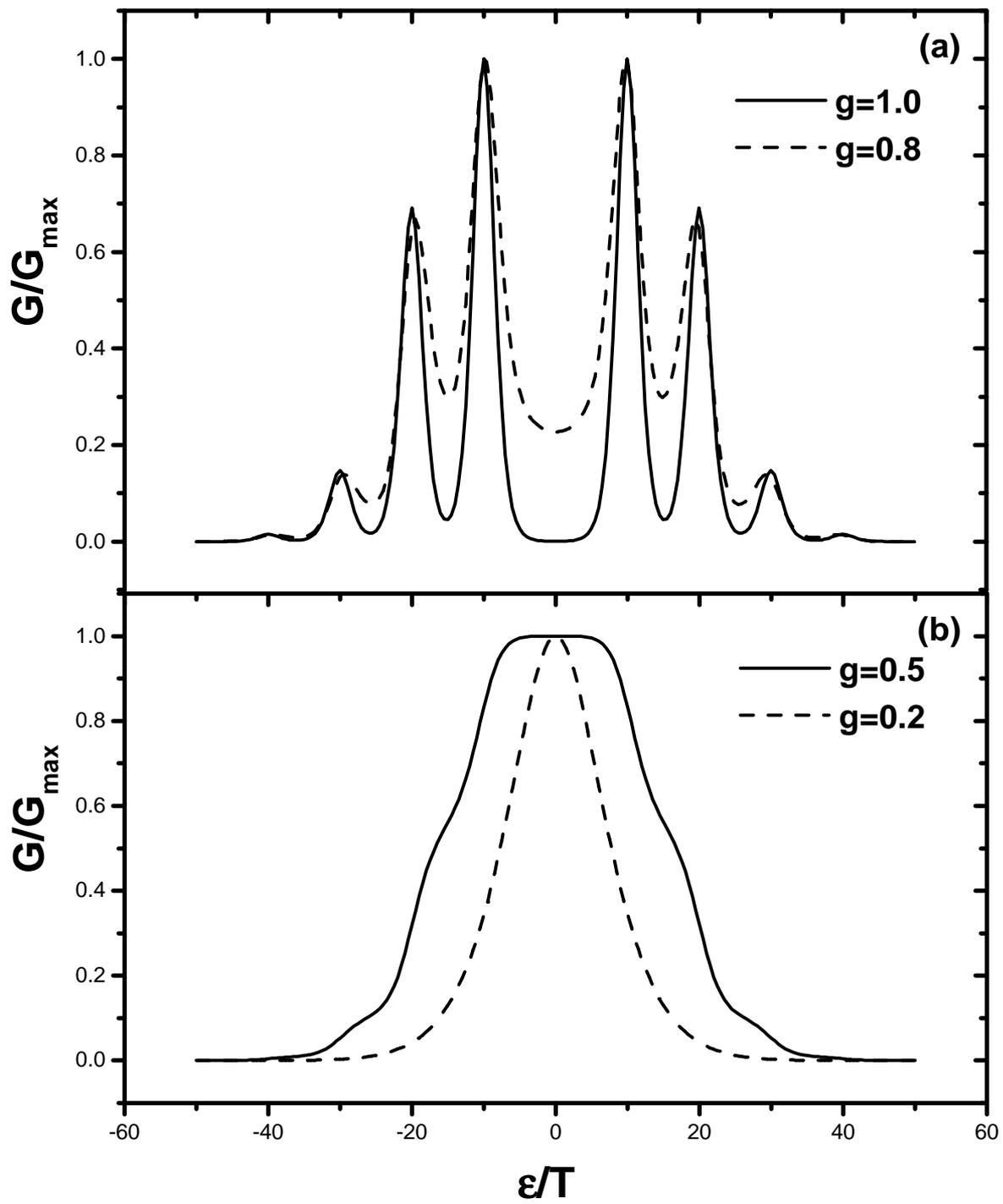
\item[{\bf Fig. 2}] $G/G_{\max }$ vs. $\varepsilon /T$ for (a) $g=1,0.8;$
(b) $g=0.5,0.2,$ with $\Omega /T=10$ and $\alpha $ the first positive root
of $J_{0}$.

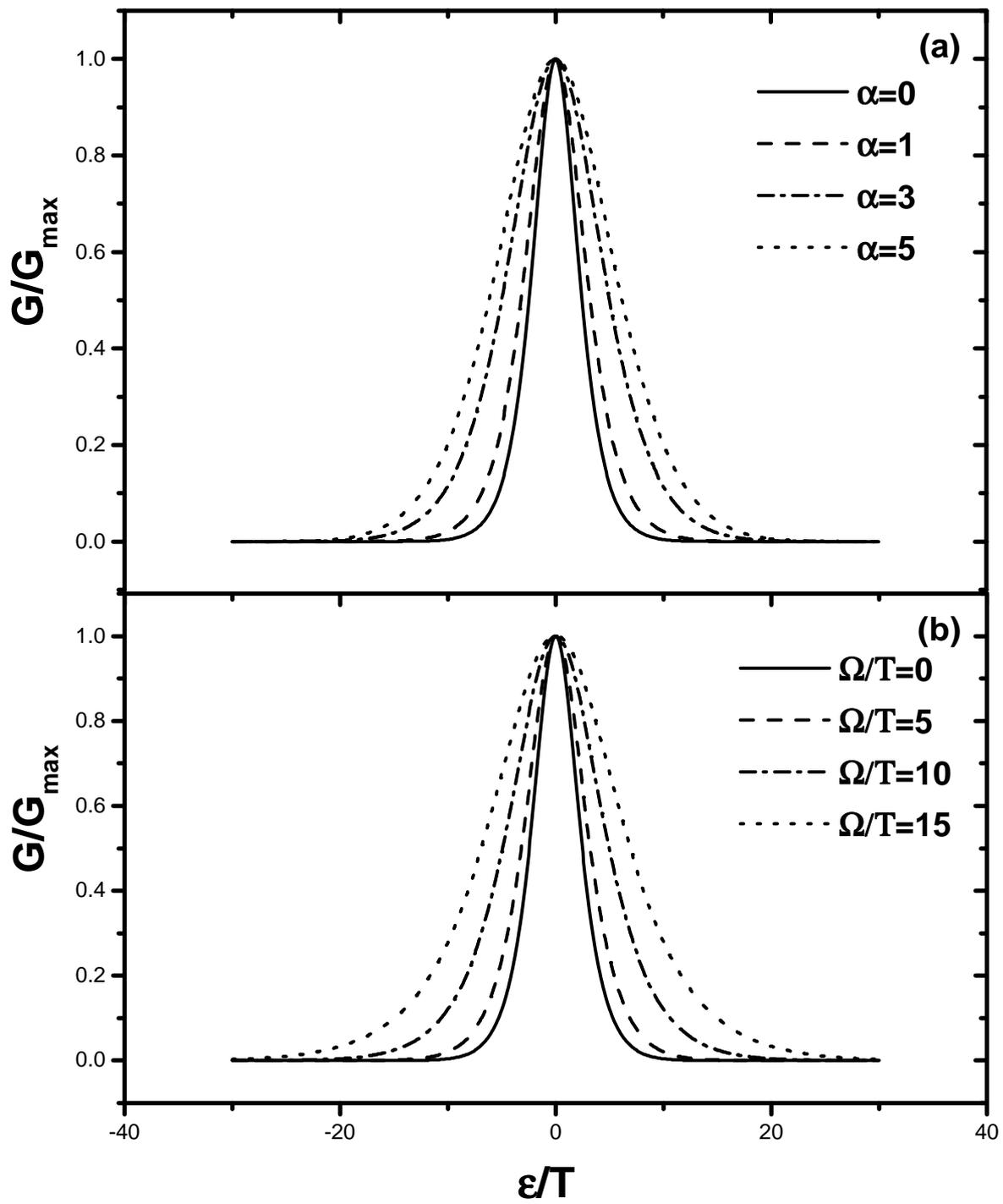
\item[{\bf Fig. 3}] $G/G_{\max }$ vs. $\varepsilon /T$ for $g=0.2$. The
parameters are chosen in (a) $\alpha =0,1,3,5$ and $\Omega /T=5$; (b) $%
\alpha =1$ and $\Omega /T=0,5,10,15.$

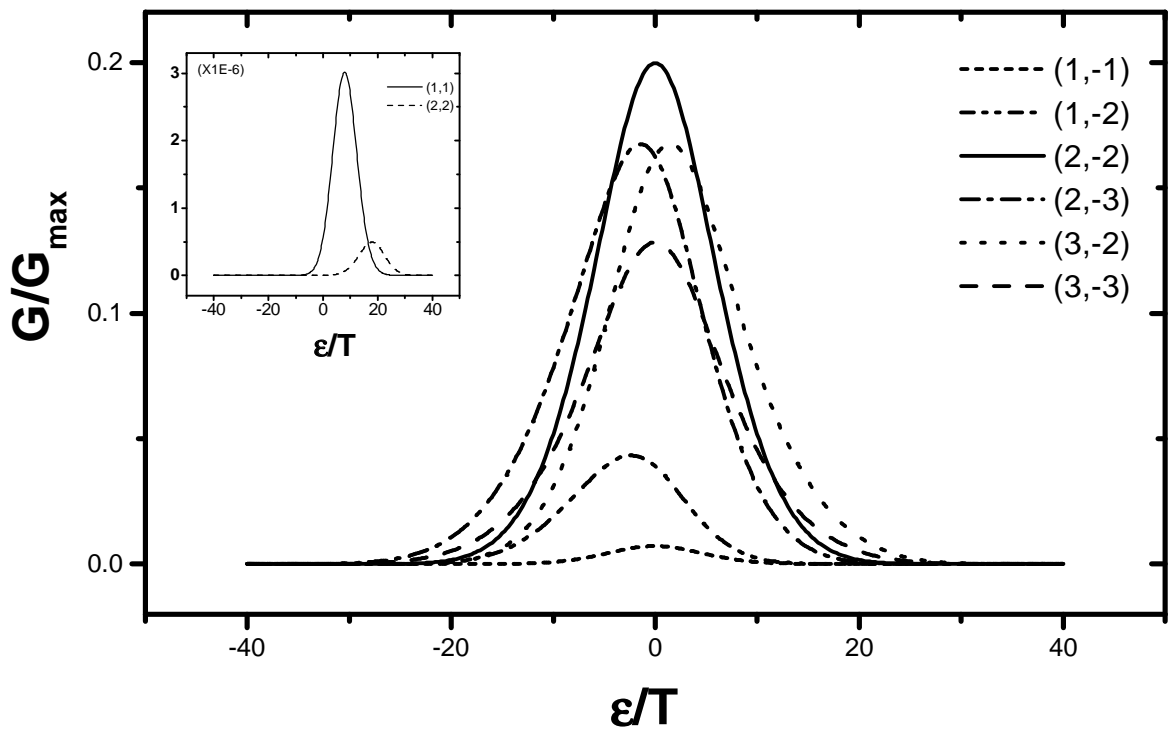
\item[{\bf Fig. 4}] The contributions of some $\left( m,n\right) $ processes
for $g=0.2,$ $\Omega /T=10,$ and $\alpha $ the first positive root of $J_{0}$%
. $G_{\max }$ is the maximum value of the total conductance.
\end{itemize}

\end{document}